\documentclass[twocolumn,aps,floatfix]{revtex4}
\usepackage{amssymb}
\usepackage{graphicx}
\usepackage{amsmath}
\usepackage[T1]{fontenc}
\usepackage{pstricks}
\usepackage{subfigure}

\begin{document}
\title{Correspondence between dark solitons and the type II excitations of Lieb-Liniger model}

\author{Tomasz Karpiuk,$^{1,2}$ Tomasz Sowi\'nski,$^{3,4}$ Mariusz Gajda,$^{3,4}$ Kazimierz Rz\k a\.zewski,$^4$ and 
Miros{\l}aw Brewczyk$\,^{1,4}$}

\affiliation{\mbox{$^1$Wydzia{\l} Fizyki, Uniwersytet w Bia{\l}ymstoku, 
                       ul. Lipowa 41, 15-424 Bia{\l}ystok, Poland}  \\
\mbox{$^2$Centre for Quantum Technologies, National University of Singapore, 3 Science Drive 2, Singapore 117543, Singapore }  \\
\mbox{$^3$Institute of Physics PAN, Al. Lotnik\'ow 32/46, 02-668 Warsaw, Poland}  \\
\mbox{$^4$Center for Theoretical Physics PAN, Al. Lotnik\'ow 32/46, 02-668 Warsaw, Poland}  }

\date{\today}

\begin{abstract}
A one-dimensional model of bosons with repulsive short-range interactions, solved analytically by Lieb and Liniger many years ago, predicts existence of two branches of elementary excitations. One of them represents Bogoliubov phonons, the other, as suggested by some authors, might be related to dark solitons. On the other hand, it has been already demonstrated within a framework of the classical field approximation that quasi-one-dimensional interacting Bose gas at equilibrium exhibits excitations which are phonons and dark solitons. By showing that statistical distributions of dark solitons obtained within the classical field approximation match the distributions of quasiparticles of the second kind derived from fully quantum description we demonstrate that type II excitations in the Lieb-Liniger model are, indeed, quantum solitons.

\end{abstract}

\maketitle

A relation between multiparticle quantum dynamics and its nonlinear classical field approximation is an intricate physical problem. A particularly interesting example is offered by a soluble model of particles moving on a circumference of a circle, interacting with contact potential known as the Lieb-Liniger model \cite{LiebI}. Its relation to a nonlinear single-particle dynamics has been recently investigated in \cite{Sato, Kanamoto}.

In this Letter we study thermal equilibrium properties of a one-dimensional uniform Bose gas in a weakly interacting regime. Our analysis is based  on the exactly solvable model \cite{LiebI,Yang} and is also performed within the classical field approximation \cite{CFA,review}. Our goal is to clarify a long-standing question regarding the physical meaning of the type II branch of the elementary excitations inherently built in the Lieb-Liniger model \cite{LiebII}. Hence, we consider a system of $N$ bosons on a circle interacting by repulsive delta-function potential. Surprisingly, such a system exhibits two families of excitations. As already identified by Lieb \cite{LiebII}, the main branch belongs to phonons what can be  confirmed by comparing the corresponding energy spectrum to the one obtained within the Bogoliubov approach. This branch typically occurs in a three-dimensional case \cite{3Dbox}.

There have been attempts to associate the second branch of excitations to dark solitons \cite{Kulish, Kolomeisky, Jackson} by analyzing the zero temperature dispersion relation for solitary waves. Here, we are going to establish the link between type II excitations and dark solitons by studying statistical distributions of excitations in a one-dimensional weakly interacting Bose gas at thermal equilibrium.

According to Landau \cite{Landau} the low temperature equilibrium properties of the system can be discussed in terms of a gas of quasiparticles. Since for a one-dimensional Bose gas of atoms interacting with the contact forces the two families of excitations come out of the Lieb-Liniger model \cite{LiebII}, a gas of two kinds of quasiparticles should be considered: phonons and type II excitations. These two kinds of quasiparticles coexist in a gas and interact with each other in an inelastic way.  For low temperatures a good approximation is to assume that all properties of the system can be derived from the ideal quasiparticle gas model. In particular, we are interested in the type II excitations' distribution at a given temperature. For that it is convenient to write the state of the system in a number representation as $\{n_{p_1}^{(1)}, n_{p_2}^{(2)}\}$. Here, $n_{p_1}^{(1)}$ and $n_{p_2}^{(2)}$ are the numbers of the Bogoliubov phonons and the type II excitations, respectively. Their momenta, $\{p_1,p_2\}$, according to Lieb \cite{LiebII}, are quantized and equally spaced with the values being integer multiples of $2\pi/L$, where $L$ is the size of the system. Moreover, the momenta of type II excitations are limited by $N\pi/L$. Total energy of the ideal gas of the quasiparticles is then given by: 
\begin{equation}
{\cal{E}}\{n_{p_1}^{(1)}, n_{p_2}^{(2)}\} = \sum_{p_1} n_{p_1}^{(1)} \epsilon_1(p_1) + \sum_{p_2} n_{p_2}^{(2)} \epsilon_2(p_2) \,,
\label{energyqp}
\end{equation}
where excitations spectra $\epsilon_1(p_1)$ and $\epsilon_2(p_2)$ depend on the only relevant dimensionless parameter of the Lieb-Liniger model $\gamma=mg/\hbar^2 \rho$, where $g$ is the strength of the atom-atom interactions and $\rho=N/L$ is the linear density of the system. The excitations spectra are found by solving the appropriate integral equations (see \cite{LiebII} for details).

\begin{figure}[thb] \resizebox{3.0in}{2.0in}
{\includegraphics{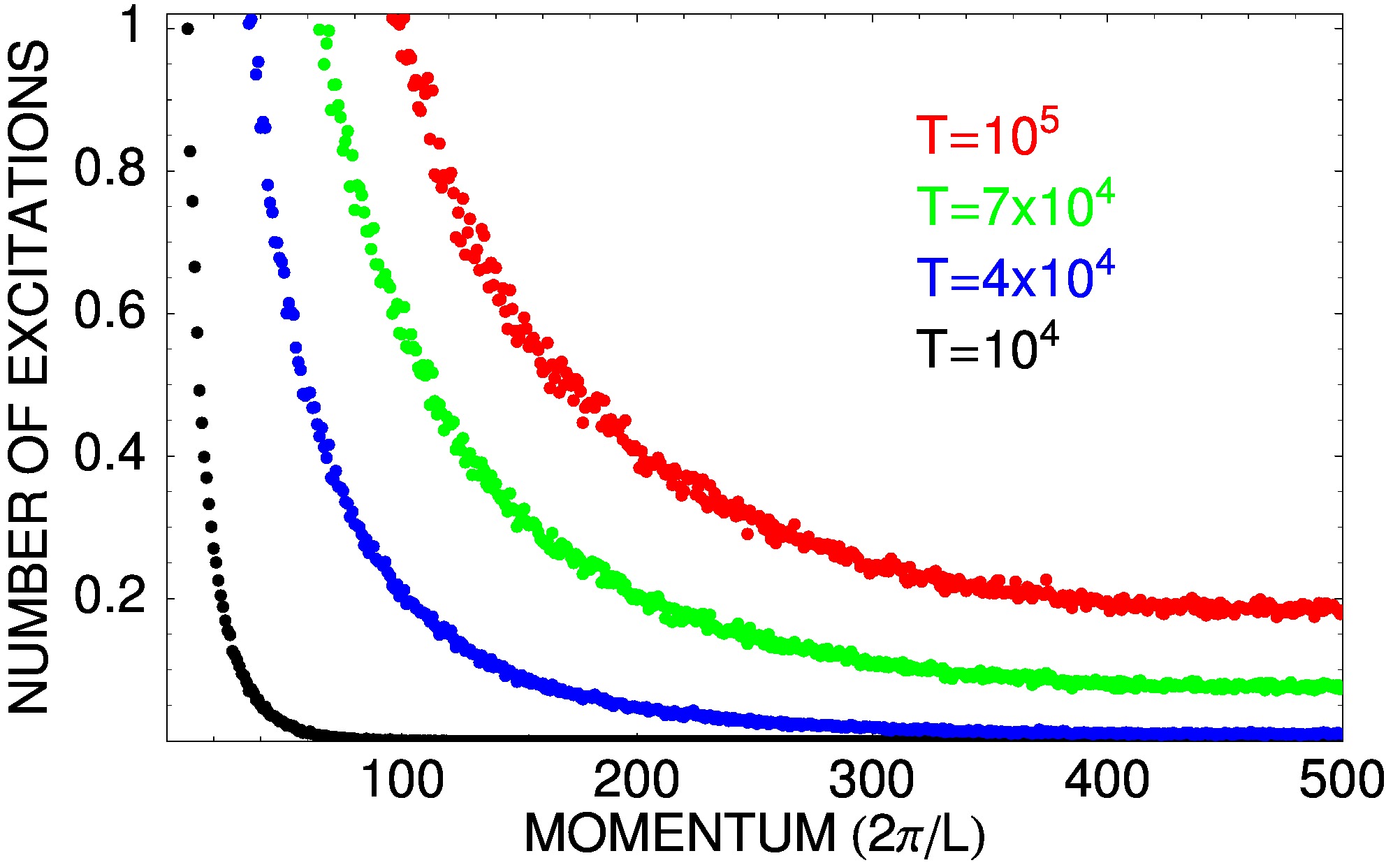}}
\caption{(color online). Average number of type II excitations as a function of quantized momentum at thermal equilibrium for the system of $N=1000$ atoms and $\gamma=0.02$. Dots correspond, from the top to bottom, to the temperatures: $10^5$, $7\times10^4$, $4\times10^4$, and $10^4$ in units of $\hbar^2/mL^2 k_B$. Maximal excitation momentum equals to $N\pi/L$. }
\label{fig1}
\end{figure}

At the thermal equilibrium, for a given temperature $T$, the states of the gas of the quasiparticles are populated according to the probability distribution
\begin{equation}
P(\{n_{p_1}^{(1)}, n_{p_2}^{(2)}\}) = \frac{1}{Z} e^{- {\cal{E}}\{n_{p_1}^{(1)}, n_{p_2}^{(2)}\}  / k_B T} \,,
\label{probabilityqp}
\end{equation}
where $Z$ is the canonical partition function. An efficient way to sample the phase space of the system of quasiparticles is to use a Monte Carlo method with the Metropolis algorithm \cite{Metropolis}. Having a canonical ensemble of states one can find various statistical averages. In particular, in Fig. \ref{fig1} we show an average number of the type II excitations for various temperatures for a system of $N=1000$ atoms and $\gamma=0.02$. We use $\hbar^2/mL^2 k_B$ as the unit of temperature. As expected, the population of higher momentum, i.e., larger energy quasiparticle states gets larger at higher temperatures.

To get the physical insight to the results presented in Fig. \ref{fig1} we now turn to the approximate treatment of an interacting Bose gas, i.e., the classical field method, \cite{review}. Within this approach, which is an extension of original Bogoliubov idea \cite{Bogoliubov}, the usual bosonic field operator $\hat {\Psi }({z})$ which annihilates an atom at point ${z}$ is replaced by the complex wave function $\Psi ({z})$. To study a thermal equilibrium state of a one-dimensional weakly interacting Bose gas we need an ensemble of classical fields $\Psi ({z})$ corresponding to a given temperature. Such an ensemble can be obtained by using the Monte Carlo algorithm \cite{Witkowska}.

In Ref. \cite{Karpiuk} we have already shown that dark solitons occur spontaneously in a one-dimensional interacting Bose gas at equilibrium. For further investigation we need a method allowing to count dark solitons. Our reasoning is as follows. Let us start with a brutally simple approximation: we assume that  the gas density is a sum of single regularized solitonic densities
\begin{equation}
\rho (x,t) = \sum_j \rho_{sol} (x-x_j-u_j t)   \,,
\label{soldensity}
\end{equation}
where $x_j$ and $u_j$ are the initial position and the velocity of $j$-th soliton, respectively. The density of a single dark soliton is given by \cite{Zakharov} \cite{Kulish}:
\begin{equation}
\rho_{sol} (x,u) = \rho_0 [ u^2-1+(1-u^2) \tanh^2(x \sqrt{1-u^2}) ]   \,.
\label{dsdensity}
\end{equation}
Here and in the rest of this paragraph we use the healing length and the speed of sound as the units of length and velocity, respectively, and $\rho_0$ is the background density. In these units $|u| \le 1$. The above approximation should be quite realistic when solitons are well separated.
Now, the Fourier transform of the density (\ref{soldensity}), both in position and time variables, results in
\begin{equation}
\tilde{\rho}_{sol} (k,\omega) =  \sum_j e^{ikx_j} f(k,u_j) \, \delta(\omega + u_j k) \,,
\label{FTdensity}
\end{equation}
where the function $f(k,u)=\int e^{ikx} \rho_{sol} (x) \, dx /2\pi$ is the Fourier transform of a single solitonic density which depends on the soliton velocity as a parameter. Since we rather use the numerical Fourier transform to calculate (\ref{FTdensity}) one should keep in mind that, in fact, the Kronecker delta appears in (\ref{FTdensity}) instead of the Dirac one. The square of the Fourier transform of the density (\ref{FTdensity}) can be calculated as
\begin{equation}
|\tilde{\rho}_{sol} (k,\omega)|^2 =  \sum_j A_j |f(k,u_j)|^2 \, \delta(\omega + u_j k) \,,
\label{FTdensity2}
\end{equation}
where the factor $A_j$ equals
\begin{equation}
A_j =  N_j + 2\sum_{j_1 > j_2} \cos(k(x_{j_1} - x_{j_2}))
\label{factorA}
\end{equation}
and $N_j$ is the number of solitons having the velocity $u_j$. To get rid of the interference terms in (\ref{factorA}) it is enough to average the formula (\ref{FTdensity2}) over many realizations since the initial positions of solitons are random. Then we have
\begin{equation}
\langle |\tilde{\rho}_{sol} (k,\omega)|^2 \rangle =  \sum_j N_j \, |f(k,u_j)|^2 \, \delta(\omega + u_j k) \,.
\label{FTdensity2ave}
\end{equation}

\begin{figure}[thb]
\begin{minipage}[b]{4cm}
\includegraphics[width=4.0cm]{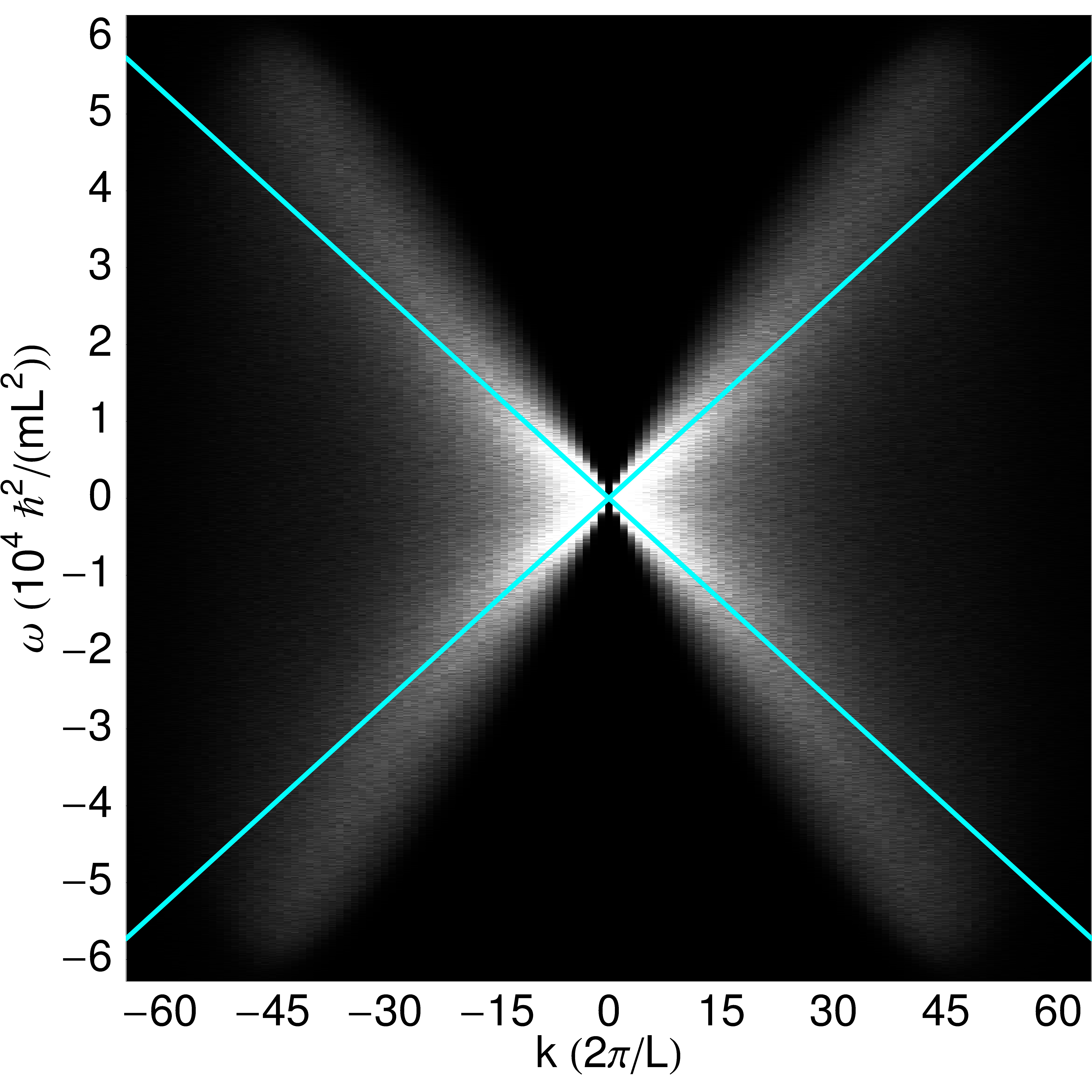}
\end{minipage}
\hfill
\begin{minipage}[b]{4.5cm}
\includegraphics[width=4.5cm,height=4.0cm]{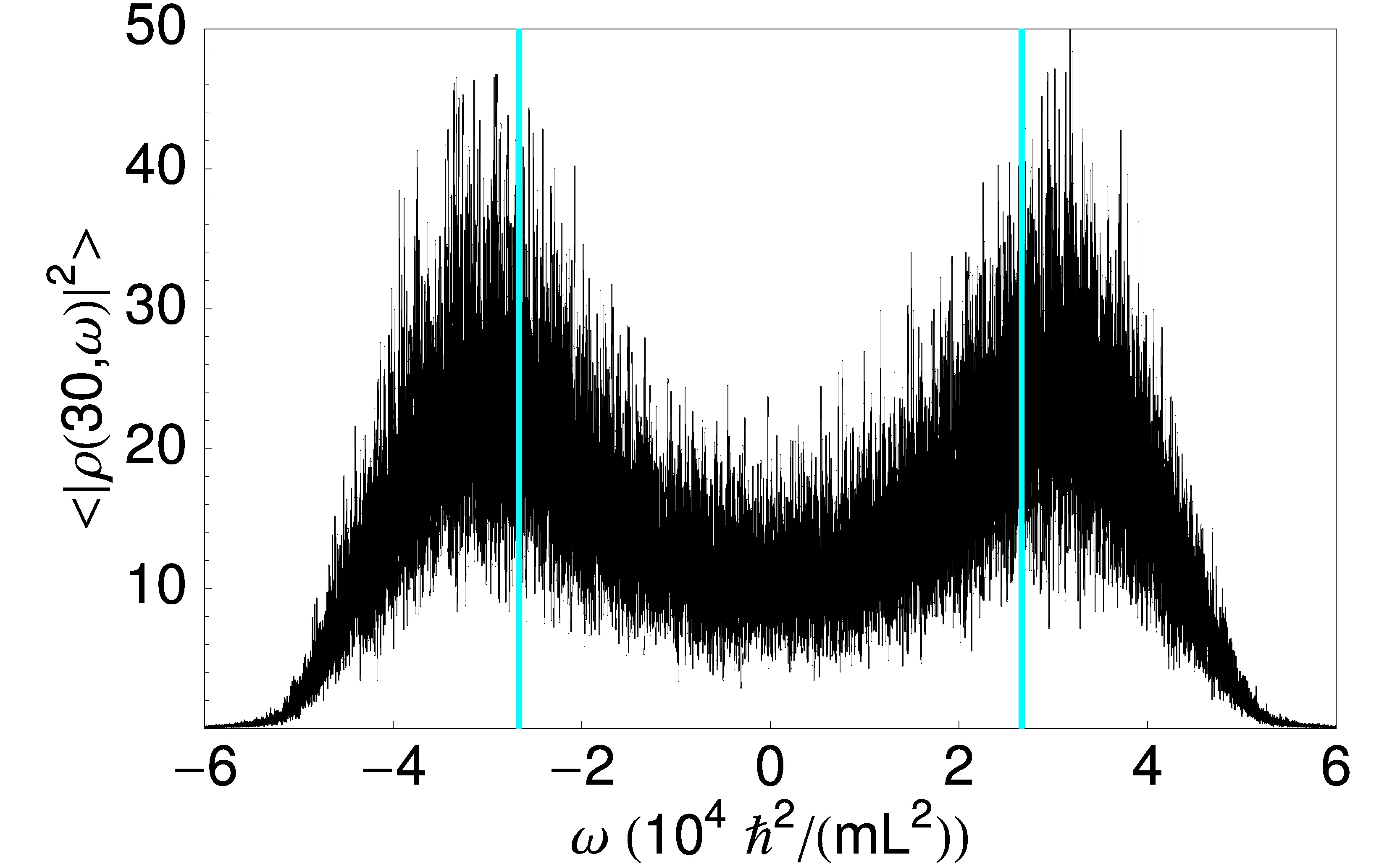}
\end{minipage}
\caption{(color online). Fourier transform, in the position and time, of the density of the classical field, first squared and next averaged over several realizations (left panel). The number of atoms $N=1000$ and the temperature is $7\times 10^4$. Blue lines are drawn according to $\omega=\pm s k$, where $s$ is the speed of sound. The right panel shows the cut for $k=30 \times 2\pi/L$. }
\label{fig2}
\end{figure}

Prescription outlined above works as follows. First, for a given temperature we prepare an ensemble of classical fields. Next, each classical field is propagated according to the Gross-Pitaevskii equation of motion. For each realization, square of the Fourier transform in position and time is calculated. Finally, an average over realizations is performed. In Fig. \ref{fig2}, left panel, we show such an average over several realizations for the system of $N=1000$ rubidium atoms at the temperature $7\times 10^4$. The superimposed blue lines are drawn as $\omega=\pm s k$, where $s$ is the speed of sound. To get an average number of dark solitons excited in a gas we need to investigate an area within the sonic cone. The right panel in Fig. \ref{fig2} shows the cut of $\langle |\tilde{\rho} (k,\omega)|^2 \rangle$ along a constant value of $k=30 \times 2\pi/L$. According to the formula (\ref{FTdensity2ave}) each piece of this data grouped around a particular value of $\omega$ carries information about the number of solitons having the velocity equal to $-\omega/k$. Therefore, dividing $\langle |\tilde{\rho} (k,\omega)|^2 \rangle$ by $|f(k,u)|^2$, where $u=-\omega/k$, gives us the number of solitons with the velocity $u$. Now, introducing a momentum of a soliton (defined as the mean value of the momentum operator calculated in the soliton state) one can relate the soliton momentum, $p(u)$, with its velocity as $p(u)=2\,(\arccos u -u \sqrt{1-u^2})$ \cite{Kulish,Jackson}. In this way we obtain the information about the spectrum of momenta of solitons which are present in a gas at thermal equilibrium. Finally, remembering that according to the Lieb model \cite{LiebII} available momenta are quantized, we count solitons by grouping their momenta around these equally spaced quantized values.

\begin{figure}[thb] \resizebox{3.0in}{2.0in}
{\includegraphics{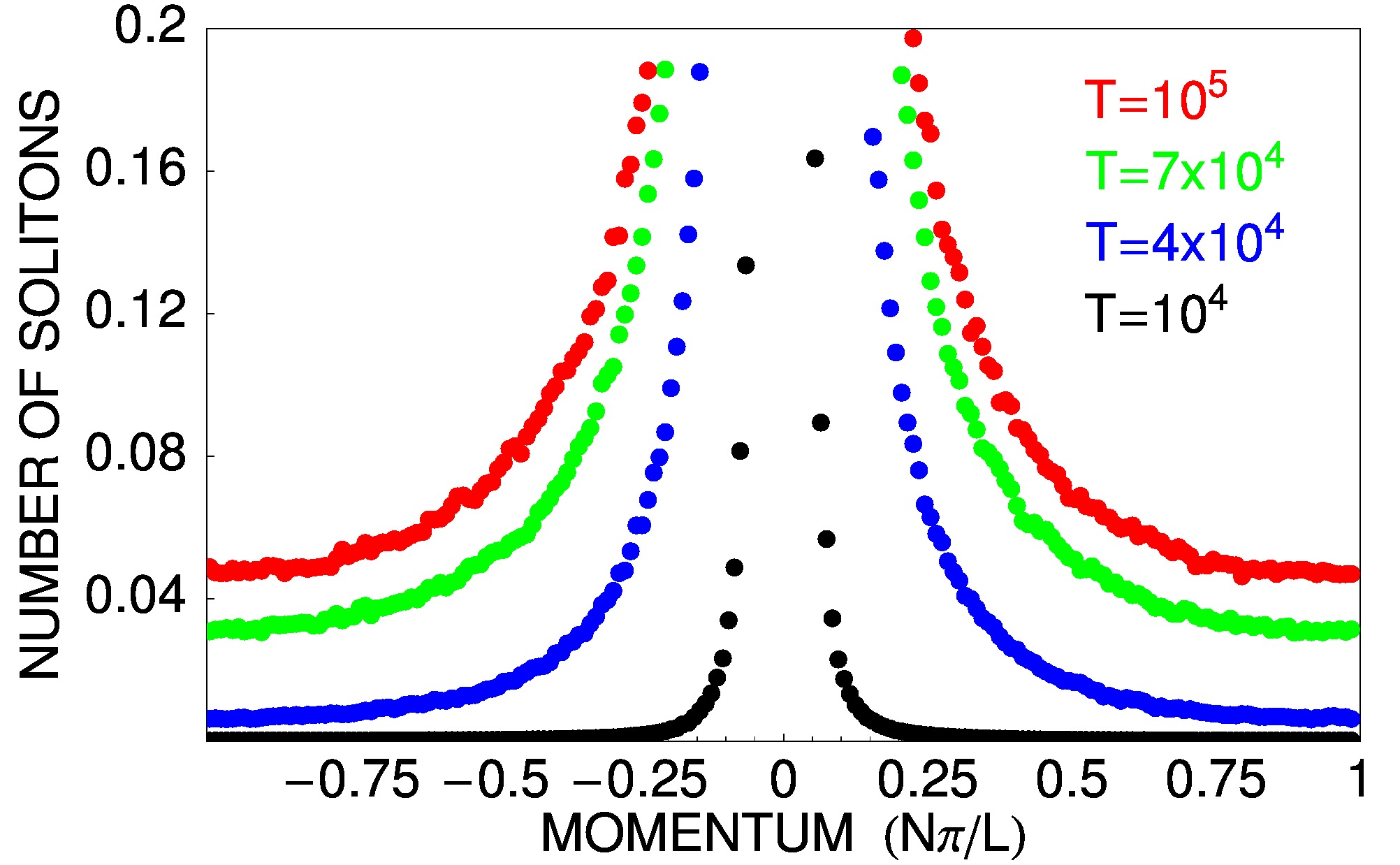}}
\caption{(color online). Average number of dark solitons as a function of momentum (in units of $\pi N/L$) at the thermal equilibrium for the system consisting of $N=1000$ atoms. Dots correspond, from the top to bottom, to the temperatures: $10^5$, $7\times10^4$, $4\times10^4$, and $10^4$.}
\label{fig3}
\end{figure}

\begin{figure}[thb]
\includegraphics[width=4.1cm]{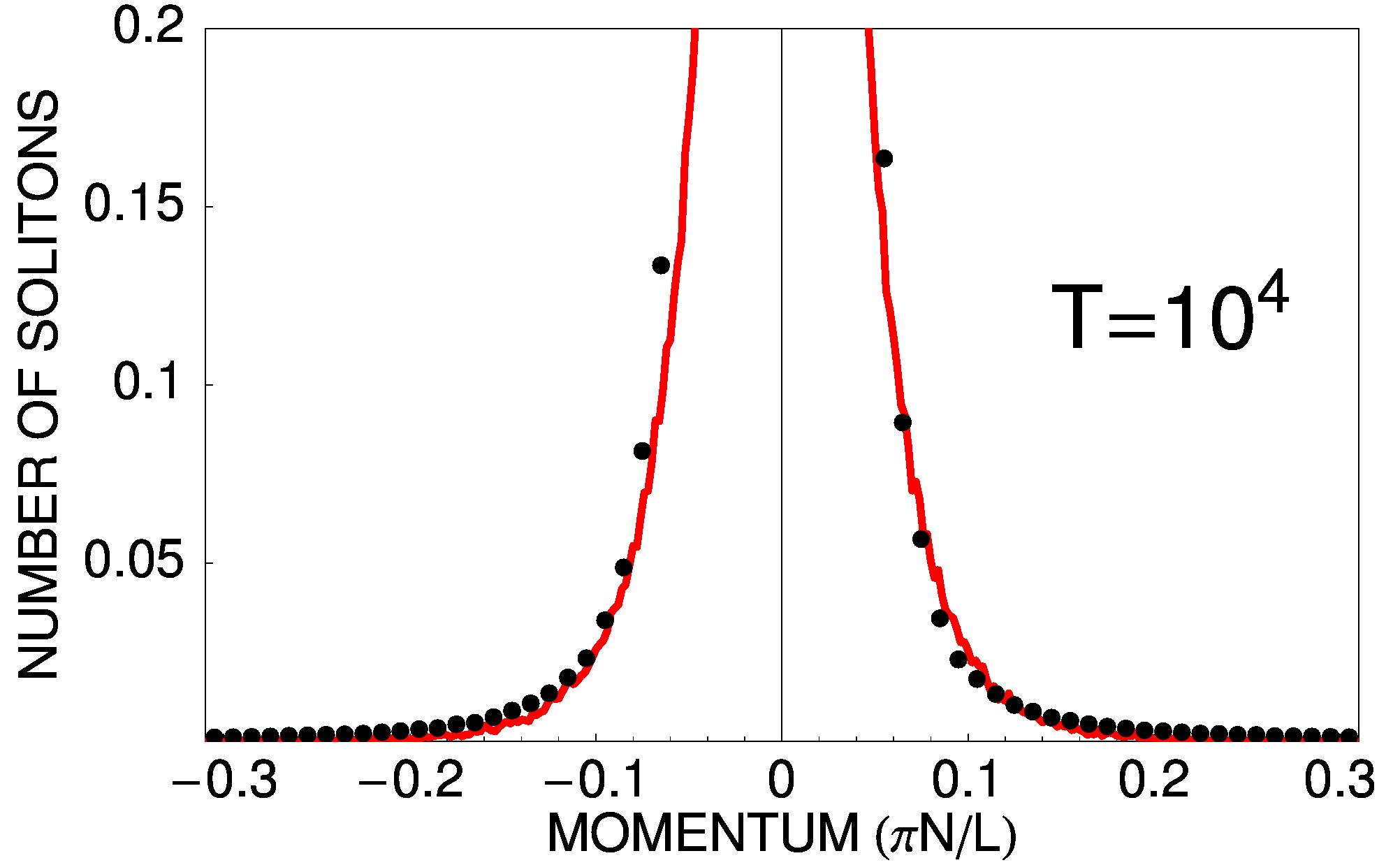}
\includegraphics[width=4.1cm]{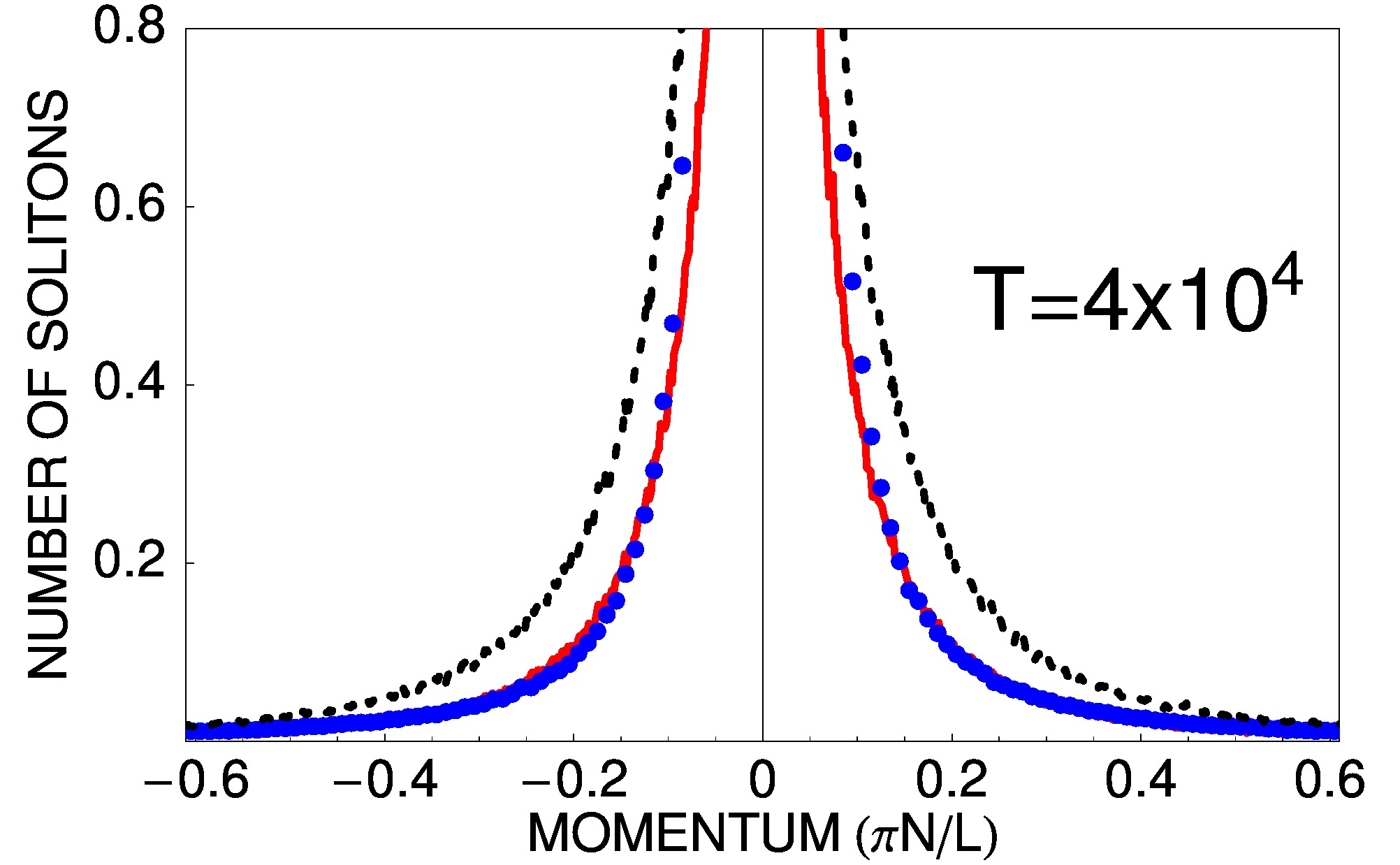}  \\
\includegraphics[width=4.1cm]{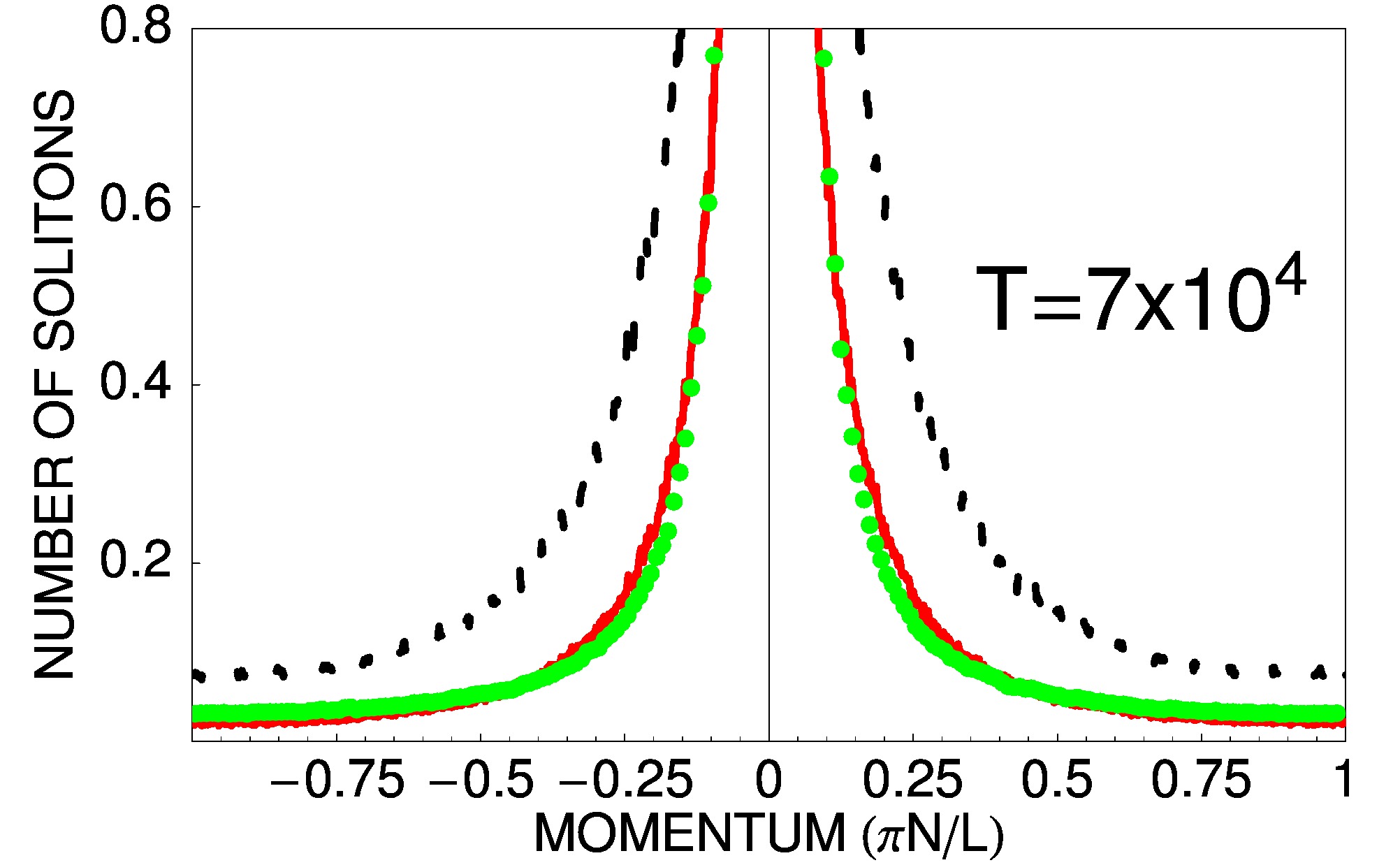}
\includegraphics[width=4.1cm]{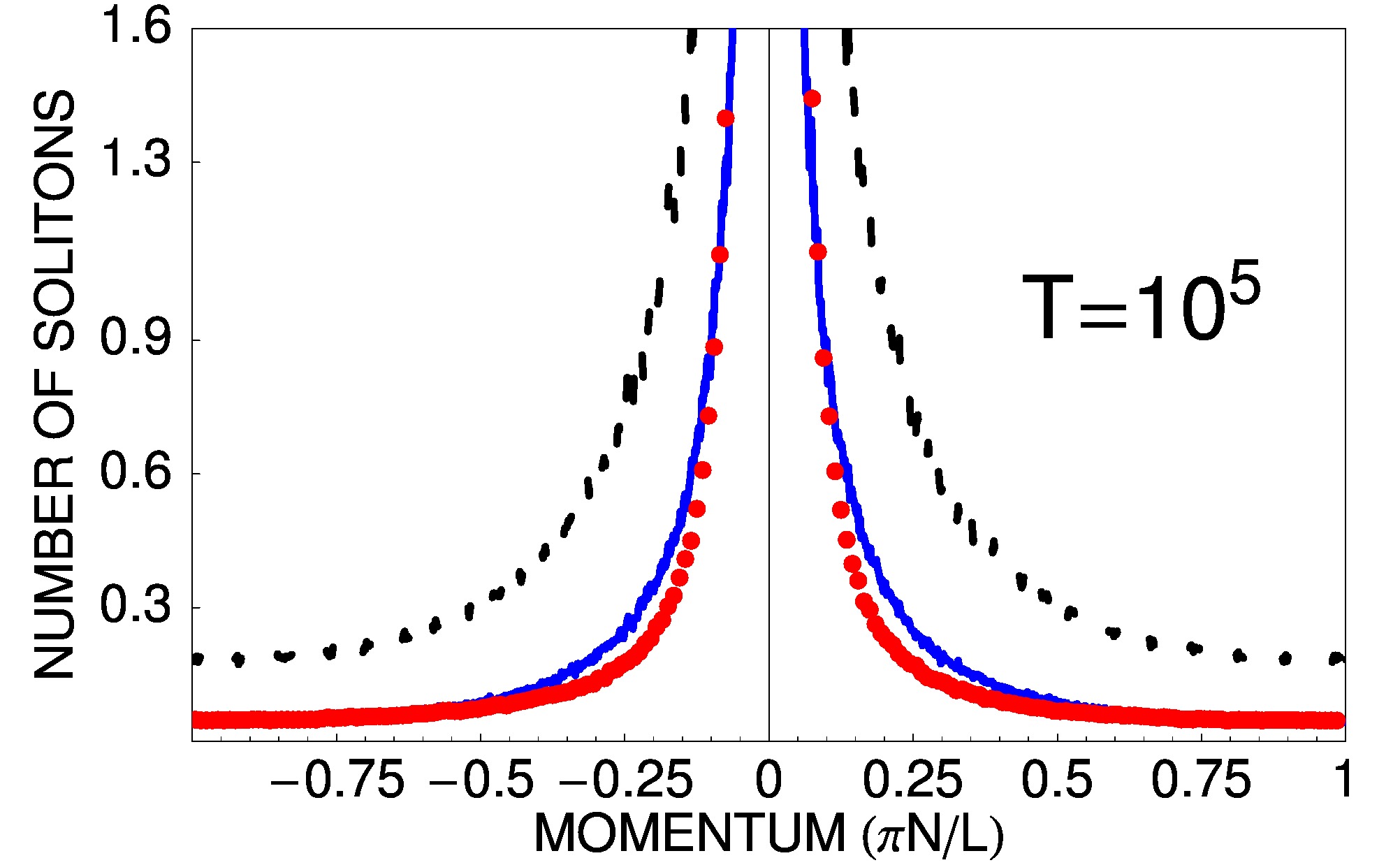} 
\caption{(color online). Comparison between the Lieb-Liniger, Yang-Yang models and the classical field approximation for four different temperatures: $T=10^4$, $4\times10^4$, $7\times10^4$, and $10^5$. Frames show the average number of the type II excitations obtained from the analytical models (solid and dashed lines for the Yang-Yang and Lieb-Liniger models, respectively) and the average number of dark solitons as calculated within the classical field approach (dots) versus the momentum. For the lowest considered temperature, $T=10^4$, the Lieb-Liniger and Yang-Yang models produce nondistinguishable curves. Note the agreement between numerical and analytical from the Yang-Yang model results. }
\label{fig4}
\end{figure}

We show the average number of dark solitons as a function of momentum in Fig. \ref{fig3} for the system consisting of $N=1000$ weakly interacting ($\gamma=0.02$) atoms. Temperatures equal to $10^4$, $4\times10^4$, $7\times10^4$, and $10^5$ are considered. As already observed in \cite{Karpiuk} deepest solitons appear for temperatures of the order of $10^5$. It is, indeed, appealing to compare our results for dark solitons with those for the Lieb-Liniger model type II excitations.  It is done in Fig. \ref{fig4}. For low temperatures (upper left frame) the agreement is very good. This is a strong indication that the Lieb-Liniger type II modes might be the quantum solitons.

Unfortunately, an agreement is spoiled for higher temperatures (see lower frames in Fig. \ref{fig4}). This can be easily understood based on the notion of the elementary excitations. For higher temperatures the model of noninteracting quasiparticles does not work and should be replaced by the one including interactions between quasiparticles. This is obviously not an easy task. However, another soluble model was developed in the late $60$s which deals with the equilibrium thermodynamics of a one-dimensional system of bosons interacting via delta function \cite{Yang}. The following set of equations (units and symbols as in the original paper \cite{Yang} are used through this paragraph)

\begin{eqnarray}
&&\epsilon(k) = -\mu +k^2 - \frac{T c}{\pi} \int_{-\infty}^{\infty} \frac{dq}{c^2+(k-q)^2}
\nonumber  \\
&&\times \ln{\{1+\exp{[\epsilon(q)/T]}\}}  \label{YY1} \\
&&2\pi \rho(k) [1+\exp{\epsilon(k)/T}] = 1+2c \int_{-\infty}^{\infty} \frac{\rho(q) dq}{c^2+(k-q)^2} 
\nonumber \\  \label{YY2}  \\
&& \frac{N}{L} = \int_{-\infty}^{\infty} \rho(k) dk    
\label{YY3}
\end{eqnarray}
allows to find the excitation spectrum at any temperature. The integral equation (\ref{YY1}) has to be  solved for a given temperature $T$, the strength of the interaction $c$, and the chemical potential $\mu$. The trial excitation spectrum $\epsilon(k)$ is used to get the density of excitations $\rho(k)$ from another integral equation (\ref{YY2}). Finally, the condition (\ref{YY3}) is checked. This is a selfconsistent procedure resulting in the spectrum $\epsilon(k)$ which, as the authors of Ref. \cite{Yang} suggest, might be regarded as the dispersion curve for a kind of effective, temperature dependent, elementary excitations.

\begin{figure}[thb] \resizebox{3.0in}{2.0in}
{\includegraphics{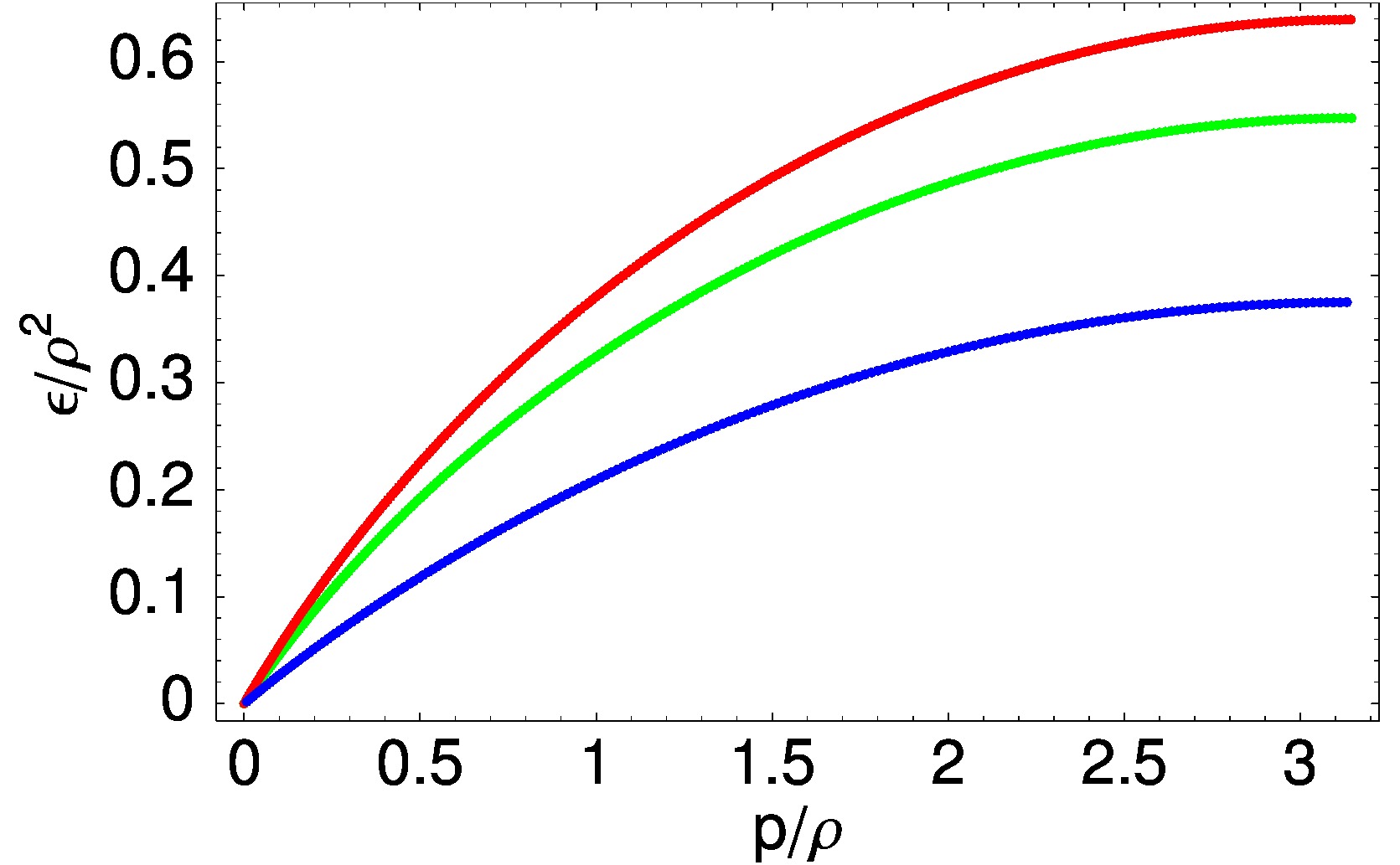}}
\caption{(color online). Dispersion curves for effective elementary excitations of the type II as obtained from Eqs. (\ref{YY1}), (\ref{YY2}), and (\ref{YY3}). Curves, from the top to bottom, correspond to temperatures $10^5$, $7\times10^4$, and zero. Here, the units of energy and momentum are $\hbar^2/(2 mL^2)$ and $N/L$, respectively. }
\label{fig5}
\end{figure}

Indeed, our results confirm this supposition. In the limit of zero temperature Eqs. (\ref{YY1}) and (\ref{YY2}) reduce to the ones obtained earlier by Lieb \cite{LiebII}. The energy curve $\epsilon(k)$ is a monotonically increasing function crossing zero value at a point which is the maximal momentum for the Lieb excitations of type II. Part of excitation curve  $\epsilon(k)$ for momenta smaller than this maximal value describes the energy of second branch of excitations. The remaining part determines the dispersion relation of the modes of the first kind. This is the procedure according to which the temperature dependent type II excitation curves were plotted, see Fig. \ref{fig5}.

Having spectra for effective temperature dependent excitations of both kinds we can now repeat the Monte Carlo calculations for the ideal gas of quasiparticles as we have already done for the Lieb-Liniger model, using Eqs. (\ref{energyqp}) and (\ref{probabilityqp}). A comparison between results obtained analytically based on the Yang-Yang model \cite{Yang} and numerically, within the classical field approximation \cite{review}, is shown in Fig. \ref{fig4}. A good agreement is clear. It means that the suggestion made by Yang and Yang \cite{Yang} regarding the excitation spectrum they found for any temperature is true. Indeed, the effective, i.e., noninteracting quasiparticles whose energies depend on temperature can be introduced in this way. But this agreement also confirms that type II excitations are solitons, in fact quantum solitons.

In summary, we have studied a uniform one-dimensional weakly interacting Bose gas at a thermal equilibrium. Properties of such a system can be investigated in terms of a gas of two families of quasiparticles introduced originally by Lieb \cite{LiebII}. For low temperatures the ideal gas approximation works well. For higher temperatures, however, quasiparticles become strongly interacting and new noninteracting effective excitations should be used according to the Yang-Yang prescription \cite{Yang}. In this way we found statistical distributions of excitations of the second type for nonzero temperatures. On the other hand, employing the classical field approximation to a uniform interacting Bose gas at thermal equilibrium we determined an average number of dark solitons for a given momentum. Both distributions match which proves that type II excitations are quantum solitons.

\acknowledgments 
%We are grateful to ... for helpful discussions. 
The work was supported by the National Science Center grants No. DEC-2011/01/B/ST2/05125 (T.K., M.G.) and DEC-2012/04/A/ST2/00090 (M.B., T.S., K.R.). The CQT is a Research Centre of Excellence funded by the Ministry of Education and the National Research Foundation of Singapore.

\end{document}